%% file: faith.tex
\documentclass[runningheads]{llncs}
\usepackage[T1]{fontenc}
\usepackage{marvosym}
\usepackage{graphicx}
\usepackage{xspace}
\usepackage{url}
\graphicspath{{./images/}}
\usepackage[hidelinks]{hyperref}
\usepackage{multirow}
\usepackage{multicol}
\usepackage{balance}
\usepackage{verbatim}
\usepackage{cancel}
\usepackage{subcaption}

\newcommand{\phrase}[1]{\textit{``#1''}}

\newcommand{\explaignn}{\textsc{Explaignn}\xspace}

\newcommand{\timequestions}{\textsc{TimeQuestions}\xspace}

\newcommand{\clocq}{\textsc{Clocq}\xspace}

\newcommand{\tsf}{{TSF}\xspace}

\newcommand{\faith}{\textsc{Faith}\xspace}
\newcommand{\benchmark}{\textsc{Tiq}\xspace}

\begin{document}
\title{A Demonstration System for Faithful Temporal Question Answering}
%
%
\author{Zhen Jia
\textsuperscript{(\Letter)} \and
Xinghong Hou \and 
Chenyang Jin 
}
\authorrunning{Z. Jia et al.}
%
\institute{Southwest Jiaotong University, Chengdu, China\\
\email{zjia@swjtu.edu.cn}\\
}

\maketitle              
%
\begin{abstract}
Temporal question answering (TQA) aims to answer questions that involve temporal constraints or require temporal information. 
We propose \faith, a novel TQA approach that operates over heterogeneous sources and ensures explainable answers by presenting traceable and faithful evidence.
In \faith, we design a new recursive answering mechanism transforming implicit temporal constraints into
explicit conditions.
Additionally, we visualize the system functionality to track how the answers are computed and the supporting evidence is derived. The code and video of the demonstration system are publicly available\footnote{\url{https://github.com/zhenjia2017/FAITH-DEMO}}.


\end{abstract}

\input{sections/01-introduction}
\input{sections/02-system}

\input{sections/05-demonstration}

\input{sections/06-conclusion}
\bibliographystyle{splncs04}
\bibliography{faith}

\end{document}

%% file: sections/01-introduction.tex
\section{Introduction}
\label{sec:intro}
\vspace*{-0.2cm}
Temporal question answering focuses on a specific type of question, temporal questions. The temporal questions can simply look up temporal information, such as \phrase{When was James Cagney born?} ($q_1$), be conditioned by explicit temporal expression, such as \phrase{Who won the Oscar for best actress in 1986?} ($q_2$), or be more complex, such as \phrase{Which club did Cristiano Ronaldo play for after Juventus FC
?} ($q_3$), with implicit temporal constraints. There has been substantial research on temporal QA~\cite{piryani2025s}.
Most methods are based on deep learning, solely encoding temporal constraints in the latent embedding space, struggling to understand the constraints precisely
and unable to explain their answer derivation process, which is called {\em unfaithful} QA. 
Another weakness of existing methods is that they operate only over a single information source: KG \textbf{or} text, inherently limiting the answer coverage of the QA system. 

Therefore, we propose \faith~\cite{jia2024faithful}, a novel approach to address these limitations. \faith introduces a new mechanism for faithful answering of temporal questions, capable of matching symbolically both explicit and implicit time constraints from fine-grained (date) to coarse-grained (month, year). \faith leverages
heterogeneous sources, including Wikidata KB, Wikipedia text, tables, and infoboxes, to achieve broad and reliable temporal coverage. To ensure explainable answer derivations, \faith applies \explaignn~\cite{christmann2023explainable}
that is based on graph neural networks (GNNs) and computes supporting evidence for the predicted answer. \faith provides an interface for facilitating users to interact with the system and trace throughout the entire answering process.



%% file: sections/02-system.tex
\begin{figure*} [t]
\centering
\includegraphics[width=\textwidth]
     {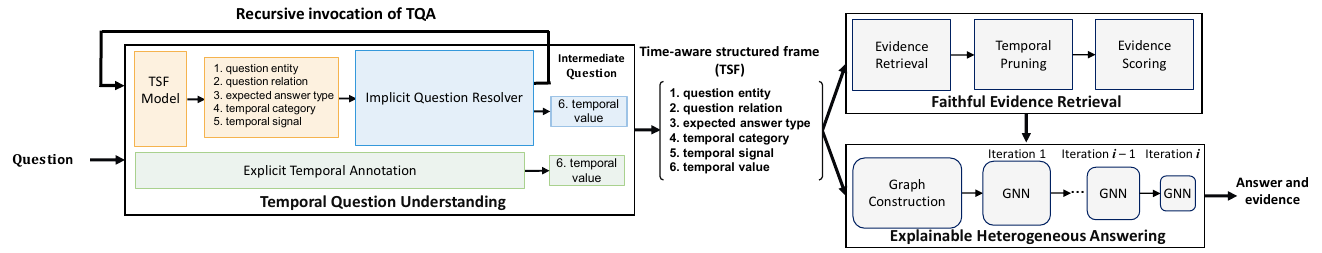}
     \vspace*{-0.3cm}
     \caption{Overview of the TQA pipeline.}
     \label{fig:tqa}
     \vspace*{-0.4cm}
\end{figure*}

\vspace*{-0.4cm}
\section{System and Key Techniques}
\vspace*{-0.2cm}
Our \faith demonstration system consists of a user interface (UI) and a backend. UI accepts users' questions with configurations and visualizes the entire TQA process, allowing users to understand how the answer and evidence are derived clearly. The backend implements the TQA approach and provides an API service. Fig.~\ref{fig:tqa} shows the key stages in the TQA approach, including: (i) temporal question understanding, (ii) faithful evidence retrieval, and (iii) explainable heterogeneous answering.

\vspace*{-0.3cm}
\subsubsection{Temporal Question Understanding}
This stage models the information needs in a structured form, \textit{\underline{T}ime-aware \underline{S}tructured \underline{F}rame (\tsf)}, and transforms implicit temporal constraints into explicit conditions. The \tsf includes six slots: \textbf{(1) question entity}, \textbf{(2) question relation}, \textbf{(3) expected answer type}, \textbf{(4) temporal category} (implicit or non-implicit), \textbf{(5) temporal signal} (e.g., before, after, or overlap), and \textbf{(6) temporal value}. We use a fine-tuned Seq2seq model, BART~\cite{lewis2020bart},
to generate the \textbf{($1$)}--\textbf{($5$)} slots in the {\tsf}. The training data for fine-tuning is obtained via
(i) distant supervision,
(ii) KB-type look-ups,
and (iii) annotations in TQA benchmarks. 
For the \textbf{temporal value}, we apply SUTime~\cite{chang2012sutime}
and regular expression matching to extract explicit temporal expressions and normalize them to a date-grained time span (e.g., the temporal value in $q_2$ is [1986-01-01, 1986-12-31]). For the implicit temporal conditions, we
proposed a novel mechanism, \textbf{implicit question resolver}, based on
\textbf{recursively invoking} the TQA system itself, to get \textbf{temporal value}. The resolver first transforms the implicit temporal constraint into \textit{intermediate question(s)}. For example, the intermediate questions
for $q_3$ would be
\phrase{When did Cristiano Ronaldo play for Juventus FC
start date?} and \phrase{When did Cristiano Ronaldo play for Juventus FC 
end date?}.
The two questions
are fed into the TQA system, respectively, to obtain two dates as answers, which are grouped into a time span [2018-01-01, 2021-08-26] as the temporal value for $q_3$. We fine-tune a BART model to generate intermediate questions. 
To create training data for fine-tuning, we utilize an LLM~\cite{ouyang2022training}
with ICL,
taking $8$ implicit questions and their corresponding intermediate questions as examples in the prompt. 

\vspace*{-0.4cm}
\subsubsection{Faithful Evidence Retrieval}
At this stage, we first use the {\tsf} as input to perform entity linking to Wikidata via \clocq~\cite{christmann2022beyond}.
With the identified entities, we retrieve relevant facts from Wikidata using \clocq again and corresponding Wikipedia pages through a locally hosted service based on the 2024-06 Wikipedia dump,
for extracting texts, tables, and infoboxes. All information is verbalized into unified textual sentences as candidate evidence. Then we use regular expression matching to annotate and normalize temporal expressions into date-grained time span, and prune out the evidence inconsistent with the temporal constraint
by the \textbf{temporal signal} and \textbf{temporal value} in {\tsf}. Finally, we use a re-ranker to score evidence and only take the top-100 evidence for computing the final answer and its explanatory
evidence. \faith integrates two re-rankers for comparison: BM25 and cross-encoder~\cite{nogueira2019passage}.
The training data for training the cross-encoder are <question, evidence> pairs with either a positive label (the evidence contains a gold answer) or a negative label (otherwise). For each question in the training set of the TQA benchmarks, we randomly sample $1$ positive pair from each knowledge source and $15$ negative pairs as the training data.

\vspace*{-0.4cm}
\subsubsection{Explainable Heterogeneous Answering}
In the final stage, we use the answering stage of \explaignn based on GNNs. We construct a heterogeneous answering graph with two types of nodes: \textbf{entities} (including dates) and \textbf{evidence}, and the links between the two types of nodes. 
The GNNs are applied iteratively to score and re-rank entities and evidence, gradually reducing the graph size to converge on the best answer while maintaining full \textbf{traceability}. We use the same methods in \explaignn to train GNN models. 
The system visualizes the reasoning steps, providing end users with clear explanations of how the answers and supporting evidence are computed. 

\vspace*{-0.4cm}
\subsubsection{Implementation}
We train \faith on two TQA benchmarks respectively: \timequestions~\cite{jia2021complex}
and \benchmark~\cite{jia2024faithful},
resulting in two suites of models. 
The pretrained LM is DistilRoBERTa\footnote{\url{https://huggingface.co/distilroberta-base}}.  
All training processes and experiments are run on a single GPU (NVIDIA Quadro RTX 8000, 48 GB GDDR6). 
\faith backend is implemented in Python, with the Flask framework providing the API service. The UI is realized in Vue.js, 
a JavaScript framework.
We utilize ECharts\footnote{\url{https://echarts.apache.org}}, an open-source JavaScript library, to visualize the charts and graphs.

%% file: sections/05-demonstration.tex
\vspace*{-0.3cm}
\section{Demonstration}

\vspace*{-0.3cm}
\subsubsection{Configuration}
Users can specify different \textbf{information sources} (e.g., KB or text) or \textbf{source combinations} (e.g., KB and text), \textbf{models} trained by \timequestions or \benchmark, \textbf{re-rankers} (BM25 or cross-encoder), if perform \textbf{temporal pruning},  \textbf{graph reduction factors} (e.g., number of iterations, number of maximum output evidence after iteration), 
enabling an analysis of how these features influence each stage in the pipeline and the final results. 
\vspace*{-0.4cm}
\subsubsection{Use case}
\begin{figure*} [t]
\centering
\includegraphics[width=\textwidth]
     {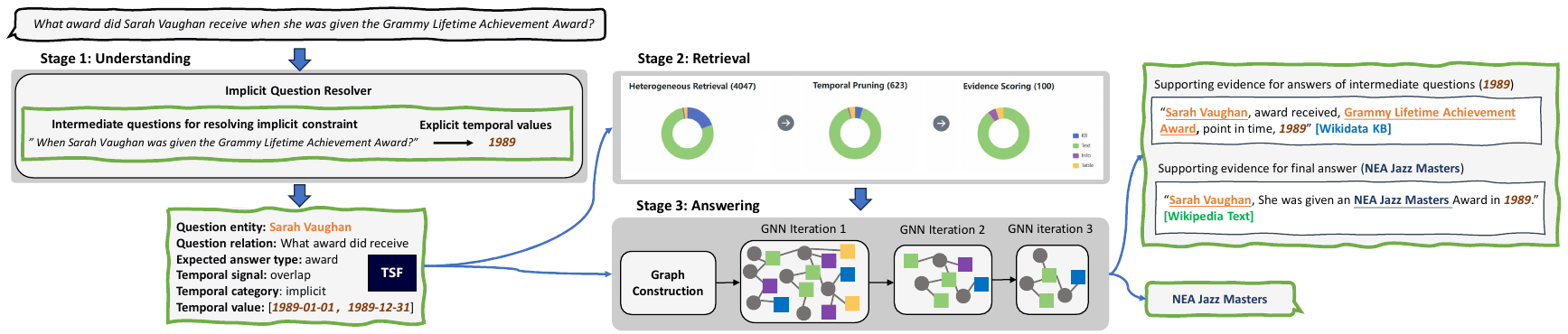}
     \vspace*{-0.3cm}
     \caption{Example demonstration.}
     \label{fig:use-case}
     \vspace*{-0.6cm}
\end{figure*}

We take a representative complex implicit question answering scenario as an example \phrase{What award did Sarah Vaughan receive when she was given the Grammy Lifetime Achievement Award?}, shown in Fig.\ref{fig:use-case}. 
In the question understanding stage, \faith generates TSF with "implicit" as its category. Therefore, the implicit question resolver generates the intermediate question \phrase{When Sarah Vaughan was given the Grammy Lifetime Achievement Award?} 
and calls the system itself to get the answer (\phrase{1989}) as the \textbf{temporal value} in TSF. In the faithful retrieval stage, \faith retrieves $4047$ evidence from sources. After temporal pruning, there are $623$ evidence left, and after scoring, the top-100 evidence are fed into the answering stage. After multiple iterations of GNN, which gradually remove the nodes in the graph, the answer, along with its supporting evidence, is computed.

%
%

%% file: sections/06-conclusion.tex
\vspace*{-0.2cm}
\section{Conclusion}
\vspace*{-0.2cm}
We created \faith, a faithful TQA system, that seamlessly combines KB, text, table and infobox, precisely understands the explicit or implicit temporal conditions, and provides a user interface for interaction and demonstrating the entire TQA process.

\begin{credits}
\vspace*{-0.2cm}
\end{credits}